# Realization of face-shear piezoelectric coefficient $d_{36}$ in PZT ceramics via ferroelastic domain engineering

Hongchen Miao[1], Faxin Li[1,2,a)]

[1]LTCS and Department of Mechanics and Engineering Science, College of Engineering, Peking University, Beijing, 100871, China

[2]HEDPS and Center for Applied Physics and Techniques, Peking University, Beijing, China

**Abstract**

The piezoelectric face-shear ($d_{36}$) mode may be the most useful shear mode in piezoelectrics, while currently this mode can only exist in single crystals of specific point groups and cut directions. Theoretically the $d_{36}$ coefficient vanishes in piezoelectric ceramics because of its transversally isotropic symmetry. In this work, we modified the symmetry of poled PZT ceramics from transversally isotropic to orthogonal through ferroelastic domain switching by applying a high lateral stress along the "2" direction and holding the stress for several hours. After removing the compression, the piezoelectric coefficient $d_{31}$ is found much larger than $d_{32}$. Then by cutting the compressed sample along the $Zxt \pm 45°$ direction, we realized $d_{36}$ coefficients up to $206 pC/N$ which is measured by using a modified $d_{33}$ meter. The obtained large $d_{36}$ coefficients in PZT ceramics could be very promising for face-shear mode resonators and shear horizontal (SH) wave generation in nondestructive testing.

---

a) Author to whom all correspondence should be addressed, Email: lifaxin@pku.edu.cn

**Introduction**

Piezoelectrics have been widely used in sensors and actuators, due to their peculiar electromechanical coupling properties, quick response and compact size.[1,2] Piezoelectric devices can be operated in a variety of deformation modes depending on the sample geometry, poling direction and field direction. The longitudinally and transversally extensional modes ($d_{33}$ and $d_{31}$) have been intensively studied and widely used. The shear deformation modes, i.e., the face-shear ($d_{36}$) mode and thickness-shear ($d_{15}$) mode, are also required in some special areas. For example, guided waves such as Lamb waves are being widely used to detect damages in plate structures.[3] However, the inherent dispersion property of Lamb wave makes the damage detection process rather complicated. An alternative solution is to utilize the non-dispersive shear horizontal (SH) waves, which can be generated by using the face-shear mode[4] or thickness-shear ($d_{15}$) mode[5] piezo-patches. However, very high operation voltage is required for the $d_{15}$ mode patches to generate the SH wave in plates, which is not suitable in practical applications.

In recent years, the discovery[6] of face shear ($d_{36}$) mode in [011]-poled PMN-PT crystals with $zxt \pm 45°$ cut direction affords a more practical face-shear mode, since its operation electric field is parallel to the poling direction (thickness direction). Experiments on relaxor-PbTiO$_3$ single crystals show that the mechanical quality factors of the face shear ($d_{36}$) mode are considerably larger than that of the thickness-shear ($d_{15}$) mode.[7,8] However, the face-shear ($d_{36}$) mode piezoelectrics has not get wide applications in the past decade.[9] The possible reason may be that currently the piezoelectric coefficient $d_{36}$ only exists in some single crystals with specific point groups and cut directions, and the low Curie temperature, poor reliability and high cost of single crystals make it difficult for industrial applications. Theoretically, the face-shear coefficient $d_{36}$ vanishes in piezoelectric ceramics because it is transversally isotropic along the poling axis.[9] Meanwhile, it is well known that piezoelectric ceramics are much more cost-effective than single crystals with better mechanical properties and higher Curie temperature. Therefore, if one can realize the face-shear ($d_{36}$) mode in piezoelectric ceramics, the related applications would surely be greatly promoted.

In this work, we changed the symmetry of poled PZT ceramics from transversally isotropic to be orthogonal by applying a high lateral stress along the "2" direction. Ferroelastic domain switching occurs and saturates during the 1$^{st}$ cycle compression with stress holding for several hours, and no further switching is observed in subsequent loading. After compression, the piezoelectric coefficient $d_{31}$ is found much larger than $d_{32}$. Then by cutting the compressed sample along the $Zxt \pm 45°$ direction, we realized large $d_{36}$ coefficients up to $206 \text{pC/N}$. The proposed $d_{36}$ generation method is repeatable and the obtained $d_{36}$ coefficients are quite stable, which may pave the way for the wide applications of face-shear piezoelectrics.

**Methods**

The PZT-5H ceramics used in this study are provided by the Institute of Acoustics, Chinese Academy of Sciences. The ceramics are cut into cube-shaped samples with dimensions of $8\text{mm} \times 8\text{mm} \times 8\text{mm}$. The piezoelectric coefficients $d_{33}$, $d_{31}/d_{32}$ of the as-received samples are measured using a precise ZJ-6A quasi-static piezo $d_{33}/d_{31}$ meter (Institute of Acoustics, Chinese Academy of Sciences). Then a compressive stress ($T_2$) perpendicular to the polar axis was applied to the samples (as shown in Fig. 1(a)) using a screw-driven testing machine (AGS-X10KN, Shimadzu Corporation, Japan) with the force resolution of $0.1\text{N}$. The maximum compressive stress is set at $155\text{MPa}$ and the loading speed is $0.05\text{MPa/s}$. When the compressive stress reached the maximum value, it was held for two hours and then removed gradually with the unloading rate of $0.5\text{MPa/s}$. After the 1$^{st}$ cycle compression, a 2$^{nd}$ cycle compression was repeated without stress holding at the maximum stress. Then the 3$^{rd}$ cycle compression was applied, which is the same as the 1$^{st}$ cycle compression, i.e., with another stress-holding for two hours. During compression loading and unloading, both the strain along the compression direction and the polarization were measured. The polarization was measured by a high-resistance ($>10^{14}\Omega$) electrometer together using a large capacitor ($1.5\mu F$) in series with the sample. The compressive strain was measured by using two $3mm \times 3mm$ strain gauges bonded on the two opposite lateral faces of the PZT sample and amplified by a 3A strain amplifier.

After 3 cycles of compression, the piezoelectric coefficients $d_{33}, d_{31}$ and $d_{32}$ are re-measured using the ZJ-6A $d_{33}/d_{31}$ meters. Then the compressed samples were cut along the $Zxt \pm 45°$ direction, as shown in Fig. 1(b)-(d). According to the IEEE Standard on piezoelectricity,[10] here the first letter $Z$ indicates the poling direction, the second letter $x$ denotes the lateral direction perpendicular to the poling direction, the third letter $t$ denotes the rotation axis which is the poling axis in this work, and the last letter $\theta$ indicates the rotation angle. After the $Zxt \pm 45°$ cut, the obtained piezoelectric sample is a tetragonal prism as shown in Fig. 1(b) and (c). The face-shear coefficient $d_{36}$ in the cut sample can then be expressed as

$$d'_{36} = 2(d_{32} - d_{31})\sin\theta\cos\theta \qquad (1)$$

Obviously $d'_{36}$ gets the maximum value at $\theta = \pm 45°$, i.e.,

$$d'_{36} = \pm(d_{32} - d_{31}) \qquad (2)$$

and

$$d'_{31} = d'_{32} = \frac{1}{2}(d_{31} + d_{32}) \qquad (3)$$

Therefore, if one can realize different values of $d_{31}$ and $d_{32}$ in piezoelectric ceramics, the $d_{36}$ can then appear in the $Zxt \pm 45°$ cut samples.

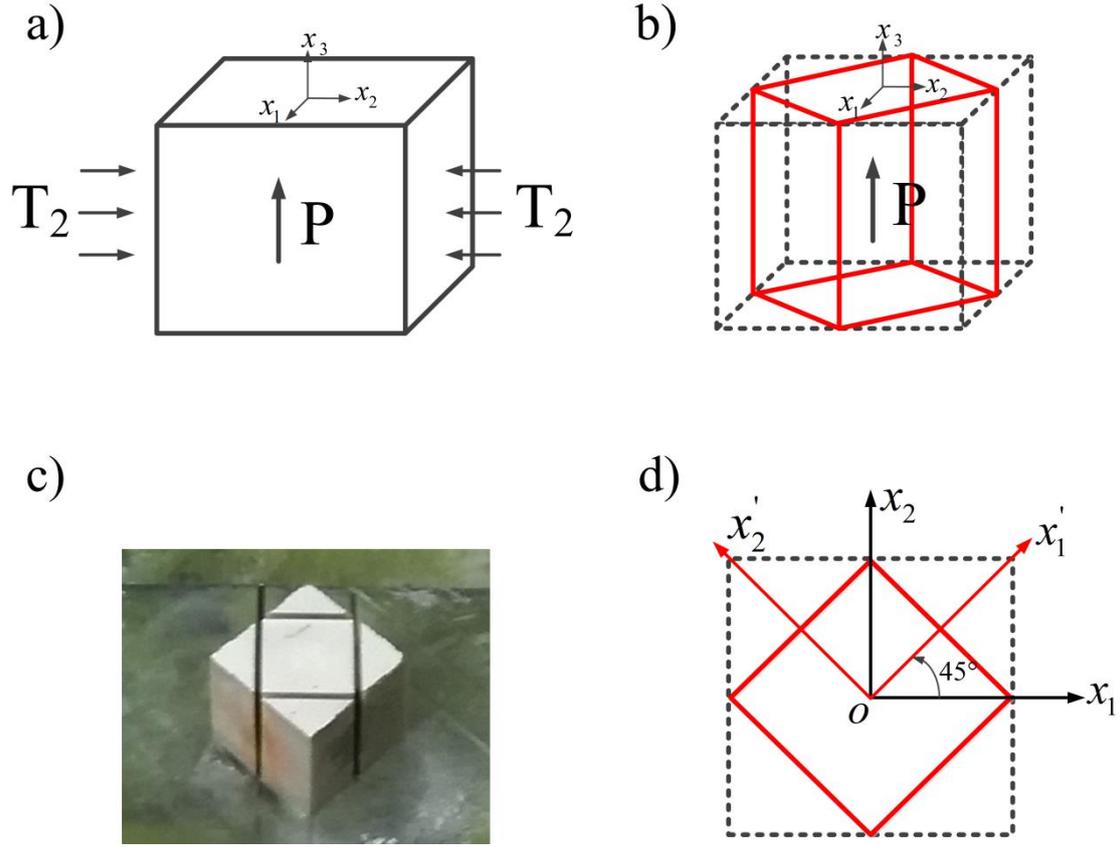

**Fig. 1.** Schematic of fabricating PZT ceramics with large piezoelectric coefficients $d_{36}$. (a) the lateral compressive stress $T_2$ is applied to a cube-shaped PZT sample to change its pseudocrystal symmetry from transversally isotropic to be orthogonal; (b) cut along the $Zxt \pm 45°$ direction, namely, the four triangular prisms (dotted lines) are cut and a tetragonal prism (red line) is obtained; (c) the photo of a $Zxt \pm 45°$ cut PZT sample; (d) top view of Fig. 1(b).

The piezoelectric coefficient $d_{36}$ in the cut samples was measured by using a charge method proposed by Weiling Yan et al..[9] The experimental setup was schematically shown in Fig. 2. The piezoelectric sample is clamped by a pair of home-made "L" shaped arms of adaptors. Then a 150Hz oscillatory force $F$ is subjected to the adaptors via the precise ZJ-6A piezo $d_{33}$ meters. Such an oscillatory force $F$ was translated to shear stress $T_6 = F/(l_2 \times l_3)$ through the adaptors, where $l_2 \times l_3$ is the area of the clamped surface. The induced charge $Q$ on the electrode is then collected by the ZJ-6A $d_{33}$ meters through wires as shown in Fig. 2. Then the electric displacement is obtained as $D_3 = Q/(l_1 \times l_2)$, where $l_1 \times l_2$ is the area of the electrode surface. Therefore, one can obtain the piezoelectric coefficients $d_{36}$ as

$$d_{36} = \frac{D_3}{T_6} = \left(\frac{Q}{F}\right) \times \left(\frac{l_3}{l_1}\right) \quad (4).$$

It should be noted that $Q/F$ is just the direct reading from the $d_{33}$ meter. Thus the measured $d_{36}$ can be expressed as

$$d_{36} = d_{33} \text{ meter reading } \times \left(\frac{l_3}{l_1}\right) \quad (5).$$

Moreover, to check the purity of the obtained face-shear mode, the impedance spectrums of sliced patches were measured using an impedance analyzer (HP4294A, Agilent Technologies) and the dynamic deformation was analyzed using the finite element method (FEM) at the resonance frequency.

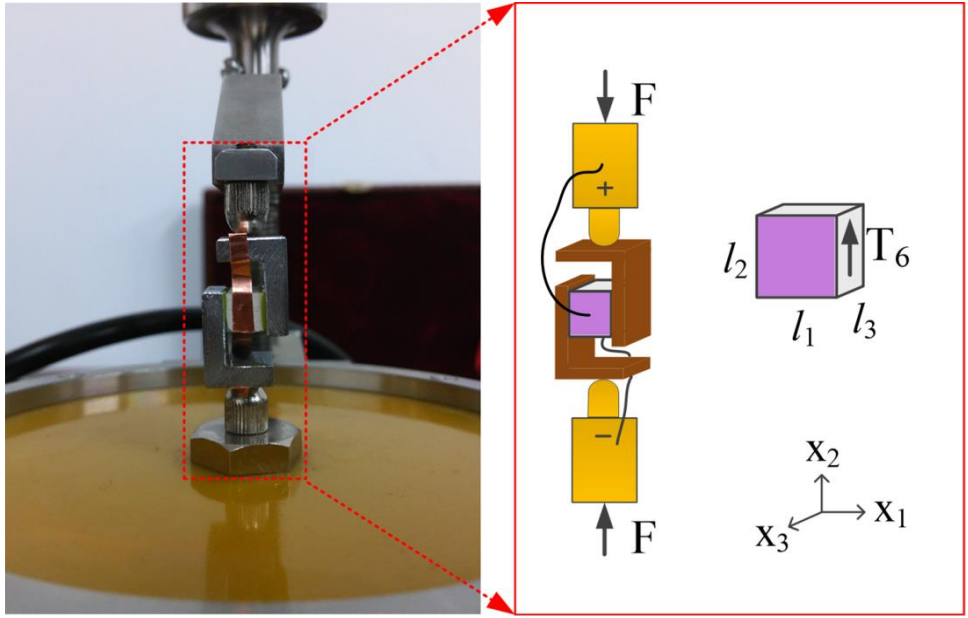

**Fig. 2.** The measurement setup for the piezoelectric coefficient $d_{36}$.

**Results and Discussions**

Fig. 3 shows the measured stress-strain curves and stress-depolarization curves of the PZT-5H ceramics during the 1st and 2nd cycles of lateral compression. It can be seen that during the 1st cycle compression, both the strain and polarization responses are significantly nonlinear which indicates that ferroelastic domain switching occurs during compression. The 1st cycle strain-stress curve in Fig. 3(a) indicates that the switchable domain is almost exhausted under the compressive stress of $85\text{MPa}$. When the stress is removed after holding the maximum stress of $155\text{MPa}$ for two hours, a remnant strain of 0.06% is obtained, which originates from the irreversible switched domains. Moreover, the unloading stress-strain curve is almost linear, indicating that little back switching occurs during unloading. In comparison, in Fig. 3(b), the polarization variations during the 1st cycle compression is quite different. It reaches the maximum value of $1.2\mu\text{C}/\text{cm}^2$ at $50\text{MPa}$, and then decreases to almost zero at the maximum stress of $155\text{MPa}$. During the two-hour stress-holding, the polarization varies from zero to about $-1\mu\text{C}/\text{cm}^2$, and it remains almost unchanged upon removing the stress. As the polarization variation is rather small compared with the remnant polarization of PZT-5H with the typical value of $41\mu\text{C}/\text{cm}^2$, the polarization variations can be regarded as negligible during lateral compression. During the 2nd cycle compression, the stress-strain curve is almost linear and the polarization keeps almost unchanged, indicating that ferroelastic domain switching had saturated and stabilized after the 1st cycle compression. The strain and polarization responses for the 3rd cycle compression are almost same as those in the 2nd cycle loading, and hence were not presented.

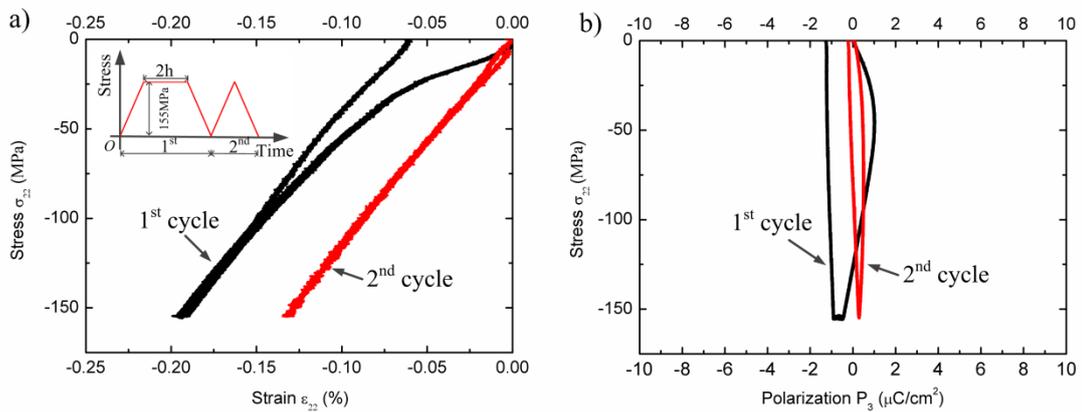

Fig. 3 Stress-strain curves (a) and stress-depolarization curves (b) of the PZT-5H ceramic during lateral compression ($T_2$).

From Fig. 3(a), it can be deduced that ferroelastic domain switching occurs during the lateral compression and the switched domains are stabilized after removing the holding compression. In this way, the symmetry of the PZT ceramics changed from transversally isotropic to orthogonal, as shown in Fig. 4. Furthermore, the ferroelastic switching induced strain from Fig. 4(a) to Fig. 4(b) can be theoretically estimated based on the saturated domain switching states in ferroelectric ceramics.[11] Take the unpoled state as the reference (zero strain), the electric poling induced strain (in Fig. 4(a)) along the $x_2$ axis is $-0.184S_0$ for the tetragonal ceramics and $-0.212S_0$ for the rhombohedral ceramics (where $S_0$ is the single-crystal deformation, $S_0 = c/a$ for tetragonal case and $S_0 = (9/8)S_{lattice}$ for rhombohedral). After a high lateral compression (in Fig. 4(b)), the saturated strain along the $x_2$ axis is $-0.269S_0$ for the tetragonal and $-0.286S_0$ for the rhombohedral. Then the ferroelastic switching induced strain along the $x_2$ axis will be $-0.085S_0$ and $-0.074S_0$ for tetragonal and rhombohedral ceramics, respectively. The PZT-5H is a morphtropic ceramics whose strain properties are more close to the rhombohedral case.[12] So take the $S_0$ for rhombohedral crystal as 0.73%[12], the compression induced ferroelastic strain is then estimated to be 0.054%, which is close to the measured value of 0.06% in Fig. 3(a).

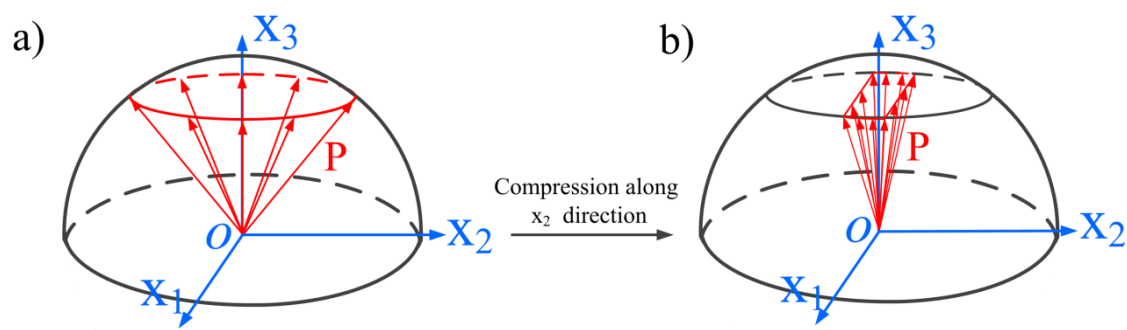

Fig. 4 Schematic of domains evolution in piezoelectric ceramics under a lateral compressive stress ($T_2$). (a) poled state along the $x_3$ axis with the transversally isotropic symmetry; (b) state after compression along the $x_2$ axis with the orthogonal symmetry.

TABLE I. Electric properties of PZT-5H samples before and after lateral compression along the $x_2$ direction.

|  | Before compression | | | After compression | | | |
| --- | --- | --- | --- | --- | --- | --- | --- |
|  | $d_{31}/d_{32}$ pC/N | $d_{33}$ pC/N | Relative dielectric constant $\kappa_{33}$ | $d_{31}$ pC/N | $d_{32}$ pC/N | $d_{33}$ pC/N | Relative dielectric constant $\kappa_{33}$ |
| Sample 1 | -297 | 692 | 4340 | -387 | -142 | 556 | 3373 |
| Sample 2 | -286 | 689 | 4343 | -372 | -154 | 555 | 3477 |
| Sample 3 | -285 | 655 | 4141 | -344 | -143 | 512 | 3248 |

Before and after the lateral compression, the electric properties of the uncut PZT samples were measured and listed in Table I. It can be seen that after compression along the $x_2$ axis, the value of $d_{32}$ decreases, while the value of $d_{31}$ increases. Such a phenomenon was also observed previously by Krueger[13] and Li et al.[14] . The change of the piezoelectric coefficients $d_{31}$ and $d_{32}$ is attributed to the ferroelastic domain switching induced crystal symmetry change from transversally isotropic to orthogonal. On the other hand, after the lateral compression, although the remnant polarization keeps almost unchanged (seen in Fig. 3(b)), both the relative dielectric constant $\kappa_{33}$ and the piezoelectric constant $d_{33}$ decreases by about 20%, which agrees with the well known relation $d = 2\kappa_0 \kappa M P_s$, where $\kappa_0$ is the permittivity of free space and $M$ is the electrostrictive coefficient. The decrease in the dielectric constant $\kappa_{33}$ is attributed to the inactive domain walls after compression[15], as a high stress can make the domains more stabilized than a large electric field[16].

From the measured $d_{31}$ and $d_{32}$ values of the uncut PZT samples in Table I, the theoretical $d_{36}'$ values of the $Zxt \pm 45°$ cut samples can be directly obtained based on Eq. (2) and are shown in Table II. The $d_{36}$ values are also measured directly using the experimental setup shown in Fig. 2. After cutting, the size of the ceramic samples turns to be $5.5mm \times 5.5mm \times 8mm$, resulting the ratio of $l_3/l_1$ to be 1.455. The measured $d_{36}$ values are also listed in Table II for comparison, from which we can see that the measured values are very close to the theoretical values for all the three samples, indicating the validity of the proposed method for generating $d_{36}$. The obtained $d_{36}$ can reach up to 206pC/N, which is comparable to the $d_{31}$ values in PZT ceramics. Furthermore, such generated $d_{36}$ is very stable and almost does not change after aging for one month.

TABLE II. Piezoelectric coefficients $d_{36}$ in the $Zxt \pm 45°$ cut PZT-5H samples.

|  | Sample 1 | Sample 2 | Sample 3 |
| --- | --- | --- | --- |
| Theoretical $d_{36}$ (pC/N) | 222 | 218 | 201 |
| Measured $d_{36}$ (pC/N) | 206 | 203 | 189 |

From Eq.(3), it can be seen that piezoelectric coefficients $d_{31}$ coexisted with $d_{36}$ in the $Zxt \pm 45°$ cut PZT samples, which is about -250pC/N for the theoretical value and about -220pC/N for the measured value. Therefore, the obtained d36 mode in PZT ceramics is not pure under quasi-static conditions, which is similar to the case in relaxor-PbTiO3 single crystals[7].

In order to check the purity of the dynamic d36 mode at the resonance frequency, sample 1 was cut into thin square patches with the dimension of $5.5mm \times 5.5mm \times 0.5mm$ and its impedance spectrum was measured by an HP4294A impedance analyzer. Fig. 5(a) shows that a clean resonance appears at 199kHz, which corresponds to the d36 face-shear mode. Furthermore, a stronger resonance peak appears at 313kHz, as shown in Fig. 5(b), which originates from the contour extensional mode due to $d_{31}$. The dynamic deformations of the square patch were then analyzed by using a finite element code (ANSYS) under a sine-wave voltage of 100V at the resonance frequencies of 199kHz and 313kHz, as shown in Fig. 5(c) and (d) where the color bar represents the Y-displacement. Obviously, the deformation at 199kHz is a pure face-shear mode, as shown in Fig. 5(c). The animation of such face-shear mode is also given in the Supplemental Material[17]. In comparison, the deformation at 313kHz is a typical extensional mode. Therefore, the pure d36 mode in the compressed PZT ceramics can be excited at its resonance frequency although it is not pure under quasi-static conditions.

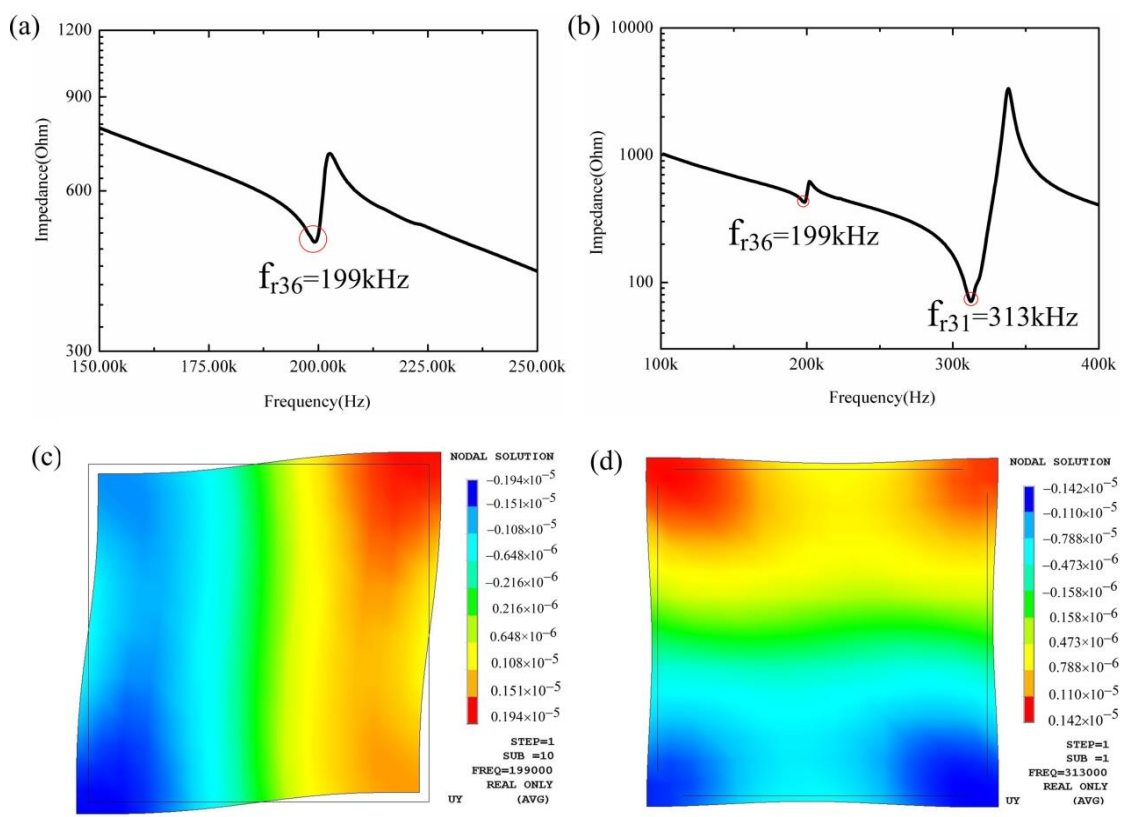

Fig. 5 (a) and (b) measured impedance as a function of frequency of a $Zxt \pm 45°$ cut PZT-5H patch; (c) and (d) FEM simulated deformation at 199kHz and 313kHz.

The obtained face-shear piezoelectric coefficient $d_{36}$ in PZT ceramics may get a variety of special applications which cannot be realized by conventional deformation modes in piezoelectrics, such as face-shear resonators, SH wave generators, etc. Although the $d_{36}$ value of PZT ceramics is still considerably smaller than that of the relaxor PMN-PT crystals ( $200 pC/N$ versus $1600 pC/N$ ), the better mechanical properties, higher Curie temperature and much lower cost of PZT ceramics could make it more competitive in industrial applications.

In summary, we proposed a reliable method to generate face-shear piezoelectric coefficient $d_{36}$ in PZT ceramics through ferroelastic domain engineering. We changed the symmetry of poled PZT ceramics from transversally isotropic to orthogonal by applying a high lateral compression and holding the stress for several hours. Then by cutting the compressed sample along the $Zxt \pm 45°$ direction, we realized face-shear piezoelectric coefficients $d_{36}$ up to $206 \text{pC/N}$. The obtained $d_{36}$ in this way is quite stable and the proposed method is applicable to all types of piezoelectric ceramics. This work will surely promote the potential wide applications based on the face-shear ($d_{36}$) mode of piezoelectrics.

HCM is grateful to Ms. Xi Chen (College of Engineering, Peking University, China) for her help in sample fabrication.